\documentclass{sig-alternate-05-2015}
\usepackage{color}
\usepackage{hyperref}
\usepackage{tikz}
\usepackage{pgfplots}
\usepackage{float}

\begin{document}

\CopyrightYear{2016}
\setcopyright{acmcopyright}
\conferenceinfo{RecSys '16,}{September 15-19, 2016, Boston , MA, USA}
\isbn{978-1-4503-4035-9/16/09}\acmPrice{\$15.00}
\doi{http://dx.doi.org/10.1145/2959100.2959160}

\clubpenalty=10000 
\widowpenalty = 10000

\title{Meta-Prod2Vec - Product Embeddings Using Side-Information for Recommendation}

\numberofauthors{3}
\author{
\alignauthor Flavian Vasile\\
\affaddr{Criteo} \\
\affaddr{Paris} \\
\email{f.vasile@criteo.com}
\alignauthor Elena Smirnova \\
\affaddr{Criteo} \\
\affaddr{Paris} \\
\email{e.smirnova@criteo.com}
\alignauthor Alexis Conneau \thanks{Alexis Conneau did this work while interning at Criteo.} \\
\affaddr{Facebook AI Research} \\
\affaddr{Paris} \\
\email{aconneau@fb.com}
}

\maketitle
\begin{abstract}
We propose Meta-Prod2vec, a novel method to compute item similarities for recommendation that leverages existing item metadata. Such scenarios are frequently encountered in applications such as content recommendation, ad targeting and web search. Our method leverages past user interactions with items and their attributes to compute low-dimensional embeddings of items. Specifically, the item metadata is injected into the model as side information to regularize the item embeddings. We show that the new item representations lead to better performance on recommendation tasks on an open music dataset.
\end{abstract}

%
%
\printccsdesc



\section{Introduction}
\label{sec:intro}

In the recent years, online commerce outpaced the growth of traditional commerce, with a rate of growth of 15\% in 2015 and accounting for \$1.5 trillion spend in 2014. The research work on recommender systems has consequently grown significantly during the last couples of years. As shown by key players in the online e-commerce space, such as Amazon, Netflix and Alibaba, the product recommendation functionality is now a key driver of demand, accounting in the case of Amazon for roughly $35\%$ \cite{amzreco} of the overall sales.

As of now, the state-of-the-art recommendation methods include matrix factorization techniques of the item-item and user-item matrices that differ in the choice of weighting schemes of the matrix entries, the reconstruction loss functions and their choice with regards to the use of additional item content information.
The real-world recommender systems have additional constraints that inform their architecture. Some of these major constraints include: scaling the recommender systems such that they can handle a huge amount of user interaction information, supporting real-time changes in recommendation \cite{chandramouli2011streamrec} and  handling the cold-start problem \cite{schein2002methods}. 
 
In the last couple of years, a promising new class of neural probabilistic models that can generate user and product embeddings has emerged and has shown promising results. The new methods can scale to millions of items and show good improvements on the cold-start problem. In the context of product recommendation, they were successfully applied to ad recommendations in Yahoo! Mail \cite{prod2vec}, for Restaurant recommendations by OpenTable \cite{opentable} and in the 1st prize winners in the 2015 RecSys Challenge \cite{romov2015recsys}.

In this paper, we present an extension of the \textit{Prod2Vec} algorithm initially proposed in \cite{prod2vec}. The \textit{Prod2Vec} algorithm only uses the local product co-occurrence information established by the product sequences to create distributed representations of products, but does not leverage their metadata. The authors have proposed an extension \cite{djuric2015hierarchical} of the algorithm that takes into account the textual content information together with the sequential structure, but the approach is specific to textual metadata and the resulting architecture is hierarchical, therefore missing some of the side information terms by comparison with our method. 
In this work, we make the connection with the work on recommendation using side information and propose \textit{Meta-Prod2Vec}, which is a general approach for adding categorical side-information to the \textit{Prod2Vec} model in a simple and efficient way. 
The usage of additional item information as side information-only, e.g. available only at training time, is motivated by real-world constraints on the number of feature values a recommender system can keep in memory for real-time scoring. In this case, using the metadata only at training time keeps the memory footprint constant (assuming an existing recommendation system that uses item embeddings) while improving the online performance.

We show that our approach significantly improves recommendation performance on a subset of the 30Music listening and playlists dataset \cite{turrin30music} with a low implementation and integration cost.

In Section \ref{sec:relatedwork} we cover previous related work and the relationship with our method. In Section \ref{sec:proposedapproach} we present the \textit{Meta-Prod2Vec} approach. In Section \ref{sec:experiments} we present the experimental setup and the results on the 30Music dataset. In Section \ref{sec:conclusion} we summarize our findings and conclude with future directions of research.

\section{Related Work}
\label{sec:relatedwork}

Existing methods for recommender systems can roughly be categorized into collaborative filtering (CF) based methods, content-based (CB) methods and hybrid methods.
CF-based methods \cite{slim} are based on user's interaction with items, such as clicks, and don't require domain knowledge. Content-based methods make use of the user or product content profiles. In practice, CF methods are more popular because they can discover interesting associations between products without requiring the heavy knowledge collection needed by the content-based methods. However, CF methods suffer from cold-start problem in which no or few interaction are available with niche or new items in the system. In recent years, more sophisticated methods, namely latent factor models, have been developed to address the data sparsity problem of CF methods which we will discuss in Section \ref{subsec:lfm}. To further help overcoming cold-start problem, recent works focused on developing hybrid methods by combining latent factor models with content information which we will cover in Section \ref{subsec:hybrid}. 

\subsection{Latent factor models}
\label{subsec:lfm}
Matrix factorization (MF) methods \cite{occf, mf} became popular after their success in the Netflix competition. These methods learn low-rank decompositions of a sparse user-item interaction matrix by minimizing the square loss over the reconstruction error. The dot product between the resulting user and item latent vectors is then used to perform recommendation. 

Several modifications have been proposed to better align MF methods with the recommendation objective, for instance, Bayesian Personalized Ranking \cite{bpr} and Logistic MF \cite{logistic-mf}. The former learns user and item latent vectors through pairwise ranking loss to emphasize the relevance-based ranking of items. The latter models the probability that a user would interact with an item by replacing the square loss in MF method with the logistic loss \cite{logistic-mf}. 

One of the first methods that learns user and item latent representations through neural network was proposed in \cite{rbm}. The authors utilized Restricted Boltzmann Machines to explain user-item interaction and perform recommendations. Recently, shallow neural networks has been gaining attention thanks to the success of word embeddings in various NLP tasks, the focus being on Word2Vec model \cite{word2vec}. An application of Word2Vec to the recommendation task was proposed in \cite{prod2vec}, called Prod2Vec model. It generates product embeddings from sequences of purchases and performs recommendation based on most similar products in the embedding space. Our work is an extension of Prod2Vec and we will present its details in Section \ref{sec:proposedapproach:prod2vec}.

\subsection{Latent factor models with content information}
\label{subsec:hybrid}
Many techniques have been used recently to create unified representations from latent factors and content information. One way to integrate user and item content information is to use it to estimate user and item latent factors through regression \cite{rlfm}. Another approach is to learn latent factors for both CF and content features, known as Factorization Machines \cite{contextfm}.

Tensor factorization have been suggested as a generalization of MF for considering additional information \cite{tensor-factorization}. In this approach, user-item-content matrix is factorized in a common latent space. The authors in \cite{metadata-co-mf} propose co-factorization approach where the latent user and item factors are shared between factorizations of user-item matrix and user and item content matrices. Similar to \cite{metadata-occf}, they also assigned weights to negative examples based on user-item content-based dissimilarity. 

Graph-based models have also been used to create unified representations. In particular, in \cite{metadata-rbfm} user-item interactions and side information are modeled jointly through user and item latent factors. User factors are shared by the user-item interaction component and the side information component. Gunawardana et al. \cite{metadata-rbm} leans the interaction weights between user actions and various features such as user and item metadata. The authors use a unified Boltzmann Machines to make a prediction.

\section{Proposed Approach}
\label{sec:proposedapproach}

\subsection{Prod2Vec}
\label{sec:proposedapproach:prod2vec}

In their \textit{Prod2Vec} paper \cite{prod2vec}, Grbovic et al. proposed the use of the \textit{Word2Vec} algorithm on sequences of product receipts coming from emails. More formally, given a set $S$ of sequences of products $s=(p_1, p_2, ..., p_M)$, $s \in S$, the objective is to find a D-dimensional real-value representation $v_p \in R^D$ such that similar products are close in the resulting vector space.

The source algorithm - \textit{Word2Vec} \cite{word2vec} is originally a highly scalable predictive model for learning word embeddings from text and belongs to the larger class of Neural Net Language Models \cite{bengio2006neural}. Most of the work in this area is based on the Distributional Hypothesis \cite{sahlgren2008distributional}, which states that words that appear in the same contexts have close if not identical meanings.

A similar hypothesis can be applied in larger contexts such as online shopping, music and media consumption and has been the basis of CF methods. In the CF setup, the users of the services are used as the distributed context in which products co-occur, leading to the classical item co-occurrence approaches in CF. A further similarity between co-count based recommendation methods and \textit{Word2Vec} has been established by Omer et al. in \cite{levy2014neural}; the authors show that the objective of the embedding method is closely related to the decomposition of the matrix containing as entries the Shifted Positive PMI of the locally co-occurring items (words), where PMI is the Point-Wise Mutual Information:

\begin{eqnarray*}
PMI_{ij}=\log\left(\frac{X_{ij} \cdot \lvert D \rvert}{X_i X_j}\right)\\
SPMI_{ij}=PMI(i,j) - \log k
\end{eqnarray*}

where $X_i$ and $X_j$ are items frequencies, $X_{ij}$ is the number of times $i$ and $j$ co-occur, $D$ is the size of the dataset and $k$ is the ratio of negatives to positives.

\paragraph{Prod2Vec Objective}
In \cite{pennington2014glove} the authors show that the \textit{Word2Vec} objective (and similarly \textit{Prod2Vec}'s) can be re-written as the optimization problem of minimizing the weighted cross entropy between the empirical and the modeled conditional distributions of context products given the target product (more precisely, this represents the Word2Vec-SkipGram model, which usually does better on large datasets). Furthermore, the prediction of the conditional distribution is modeled as a softmax over the inner product between the target product and context product vectors:

\begin{eqnarray*}L_{\text{P2V}} &=& L_{J|I}(\theta) \\
&=&  \sum_{ij} (-X^{\text{POS}}_{ij} \log q_{j|i}(\theta) - (X_i - X^{\small{\text{POS}}}_{ij}) \log (1 - q_{j|i}(\theta))\\
 &=&  \sum_{ij} X_i (- p_{j|i} \log q_{j|i}(\theta)- p_{\neg j|i} \log q_{\neg j|i}(\theta)) \\
 &=&  \sum_{i} X_i H(p_{\cdot|i}, q_{\cdot|i}(\theta)).
\end{eqnarray*}

Here, $H(p_{\cdot|i}, q_{\cdot|i}(\theta))$ is the cross-entropy of the empirical probability $p_{\cdot|i}$ of seeing any product in the output space $J$ conditioned on the input product $i \in I$ and the predicted conditional probability $q_{\cdot|i}$: 
\begin{eqnarray*}
q_{j|i}(\theta)= \frac{exp(w_i^T w_j)}{exp(w_i^T w_j) + \sum_{j' \in (V_{J-j})} exp(w_i^T w_j')}
\end{eqnarray*}
where $X_i$ is the input frequency of product i and $X^{\text{POS}}_{ij}$ is the number of times the pair of products $(i,j)$ has been observed in the training data.

\begin{figure}[t]
\begin{center}
  \includegraphics[scale=0.25]{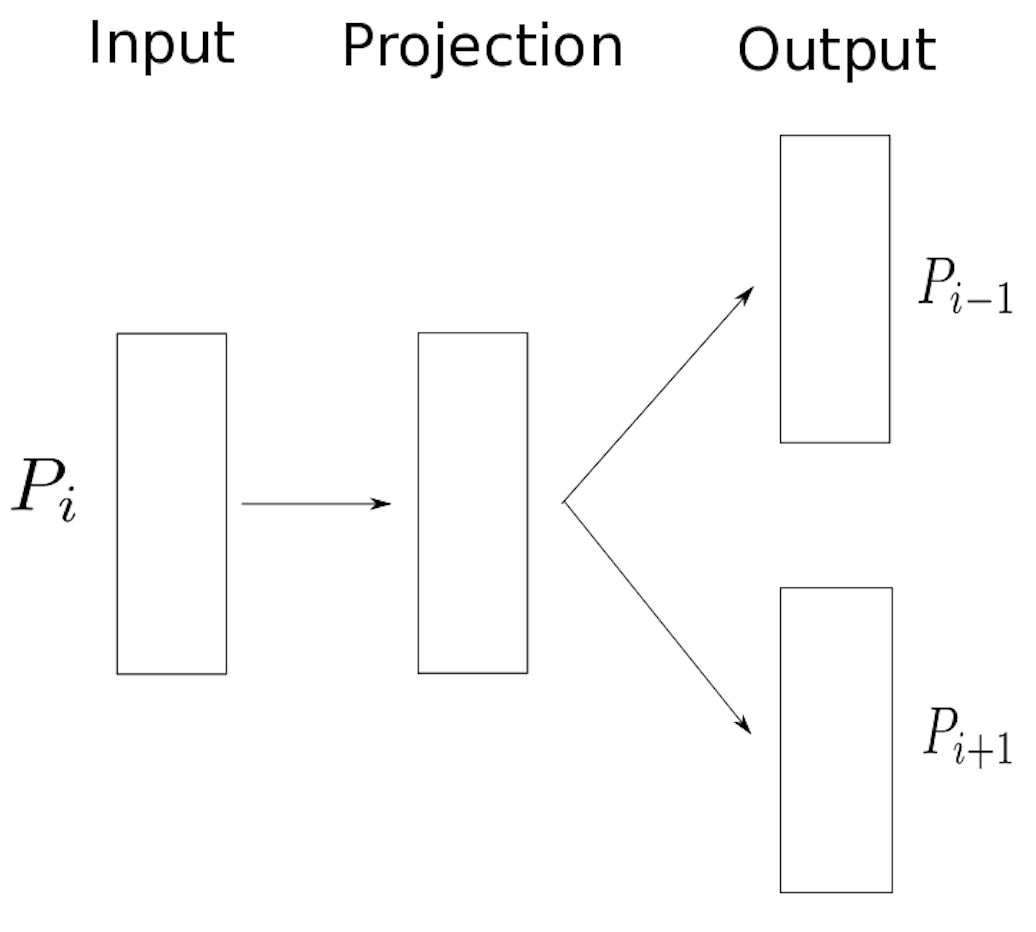}
  \caption{Prod2Vec Neural Net Architecture.}
  \label{fig:nn_prod2vec}
  \end{center}
\end{figure}

The resulting architecture for \textit{Prod2Vec} is shown in Figure \ref{fig:nn_prod2vec}, where the input space of all products situated in center window is trained to predict the values of the surrounding products by using a Neural Network with a single hidden layer and a softmax output layer.

However, the product embeddings generated by \textit{Prod2Vec} only take into account the information of the user purchase sequence, that is the local co-occurrence information. Though richer than the global co-occurrence frequency used in Collaborative Filtering, it does not take into account other types of item information that are available (the items' metadata).

For example, assuming that the inputs are sequences of categorized products, the standard \textit{Prod2Vec} embedding does not model the following interactions:
\begin{itemize}
\item given the current visit of the product $p$ with category $c$, it is more likely that the next visited product $p'$ will belong to the same category $c$
\item given the current category $c$, it is more likely that the next category is $c$ or one of the related categories $c'$ (e.g. after a swimwear sale, it is likely to observe a sunscreen sale, which belongs to an entirely different product category)
\item given the current product $p$, the next category is more likely to be $c$ or a related category $c'$
\item given the current category $c$, the more likely current products visited are $p$ or $p'$
\end{itemize}
 
As mentioned in the introduction, the same authors extended the \textit{Prod2Vec} algorithm in \cite{djuric2015hierarchical} such that to take into account both product sequence and product text in the same time. If applying the extended method to non-textual metadata, the algorithm models, additionally to the product sequence information, the dependency between the product metadata and the product id, but does not link the sequence of metadata and the sequence of product ids together.

\subsection{Meta-Prod2Vec}
\label{sec:proposedapproach:metaprod2vec}
As shown in the Related Work section, there has been extensive work on using side information for recommendation, especially in the context of hybrid methods that combine CF methods and Content-based (CB) methods. In the case of embeddings, the closest work is the \textit{Doc2Vec} \cite{le2014distributed} model where the words and the paragraph are trained jointly, but only the paragraph embedding is used for the final task.

\begin{figure}[t]
\begin{center}
  \includegraphics[scale=0.22]{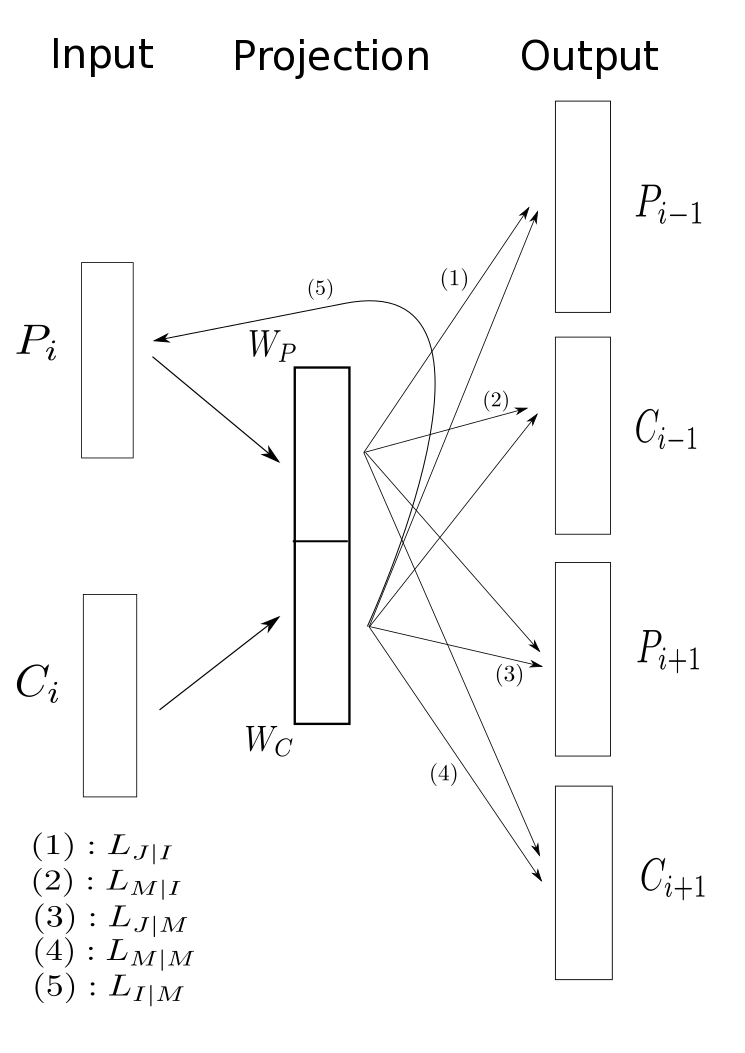}
  \caption{Meta-Prod2Vec Neural Net Architecture.}
  \label{fig:nn_metaprod2vec}
  \end{center}
\end{figure}

We propose a similar architecture that incorporates the side information in both the input and output space of the Neural Net and parametrizes separately each one of the interactions between the items to be embedded and the metadata, as shown in Figure \ref{fig:nn_metaprod2vec}.

\paragraph{Meta-Prod2Vec Objective}
The \textit{Meta-Prod2Vec} loss extends the \textit{Prod2Vec} loss by taking into account four additional interaction terms involving the items' metadata:

\begin{eqnarray*}
L_{MP2V} = L_{J|I} + \lambda \times (L_{M|I} + L_{J|M} + L_{M|M} + L_{I|M})
\end{eqnarray*}

where: M is the metadata space (for example, artist ids in the case of the 30Music dataset), $\lambda$ is the regularization parameter. We list the new interaction terms below: \\

$L_{I|M}$ is the weighted cross-entropy between the observed conditional probability of input product ids given their metadata and the predicted conditional probability. This side-information is slightly different from the next three types because it models the item as a function of its own metadata (same index in the sequence). This is because, in most cases, the item's metadata is more general than the id and can partially explain the observation of the specific id.\\

$L_{J|M}$ is the weighted cross-entropy between the observed conditional probability of surrounding product ids given the input products' metadata and the predicted conditional probability. An architecture where the normal \textit{Word2Vec} loss is augmented with only this interaction term is very close to the \textit{Doc2Vec} model proposed in \cite{mikolov2013distributed} where we replace the document id information with a more general type of item metadata.\\

$L_{M|I}$ is the weighted cross-entropy between the observed conditional probability of surrounding products' metadata values given input products and the predicted conditional probability.\\

$L_{M|M}$ is the weighted cross-entropy between the observed conditional probability of surrounding products' metadata values given input products metadata and the predicted conditional probability. This models the sequence of observed metadata and in itself represents the \textit{Word2Vec}-like embedding of the metadata.\\

\begin{figure}[t]
\begin{center}
  \includegraphics[scale=0.38]{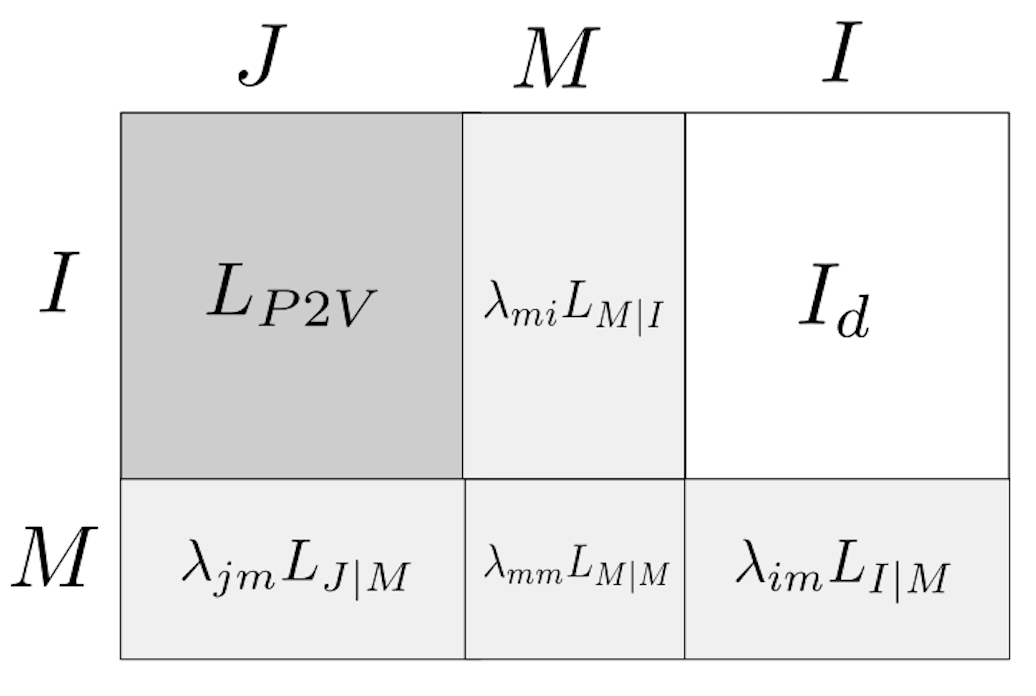}
  \caption{MetaProd2vec as matrix factorization of the items and metadata extended matrix.}
  \label{fig:matrixIMJ}
  \end{center}
\end{figure}

To summarize, $L_{J|I}$ and $L_{M|M}$ encode the loss terms coming from modeling the likelihood of the sequences of items and metadata separately, $L_{I|M}$ represents the conditional likelihood of the item id given its metadata and $L_{J|M}$ and $L_{M|I}$ represent the cross-item interaction terms between the item ids and the metadata. In Figure \ref{fig:matrixIMJ} we show the relationship between the item matrix factorized by \textit{Prod2Vec} and the one factorized by \textit{Meta-Prod2Vec}.

The more general equation for \textit{Meta-Prod2Vec} introduces a separate $\lambda$ for each of the four types of side-information: $\lambda_{mi}$, $\lambda_{jm}$, $\lambda_{mm}$ and $\lambda_{im}$. 

In Section \ref{sec:experiments} we will analyze the relative importance of each type of side-information. Also, in the case when multiple sources of metadata are used, each source will have its own term in the global loss and its own regularization parameter.

In terms of the softmax normalization factor, we have the option of either separate the output spaces of the items and of their metadata or not. Similarly with the simplifying assumption used in \textit{Word2Vec}, that allows each pair of co-occurring products to be predicted and fitted independently (therefore adding an implicit mutual exclusivity constraint on the output products given an input product), we embed the products and their metadata in the same space, therefore allowing them to share the normalization constraint.

One of the main attractions of the \textit{Word2Vec} algorithm is its scalability, which comes from approximating the original softmax loss over the space of all possible words with the Negative Sampling loss \cite{mikolov2013distributed, nce}, that fits the model only on the positive co-occurrences together with a small sample of the negative examples to maximize a modified likelihood function $L_{SG-NS}(\theta)$:

\begin{eqnarray*}
L_{J|I}(\theta) = \sum_{ij} (-X^{POS}_{ij} \log q_{j|i}(\theta) - (X^{NEG}_{ij} \log (1 - q_{j|i}(\theta)) \\ \approx L_{SG-NS}(\theta)
\end{eqnarray*}

and:

\begin{eqnarray*}
L_{SG-NS}(\theta) = \sum_{ij} -X^{POS}_{ij} (\log \sigma(w_i^T w_j) \\ - k \mathbb{E}_{c_N \sim P_D} \log \sigma(-w_i^T w_N))
\end{eqnarray*}

where $P_D$ probability distribution used to sample negative context examples and $k$ is a hyper parameter specifying the number of negative examples per positive example. The side-information loss terms $L_{I|M}$ , $L_{J|M}$, $L_{M|I}$, $L_{M|M}$ are computed according to the same formula, where the i,j indexes range over the respective input/output spaces.

In the case of \textit{Meta-Prod2Vec}, the impact of the decision to co-embed products and their metadata on the $L_{SG-NS}(\theta)$ loss is that the set of potential negative examples for any positive pair ranges over the union of items' and metadata values.

\paragraph{Impact on final recommendation system. Feasibility and engineering concerns}
Because of the shared embedding space, the training algorithm used for \textit{Prod2Vec} remains unchanged. The only difference is that, in the new version of the generation step of training pairs, the original pairs of items are supplemented with additional pairs that involve metadata. 
In terms of the online recommendation system, assuming we are augmenting a solution that already involves item embeddings, the online system does not incur any changes (since the only time we make use of the metadata is during training) and there is zero additional impact on the online memory footprint. 

\section{Experiments}
\label{sec:experiments}
The experimental section is organized as follows. First, we describe the evaluation setup, namely, the evaluation task, success metrics and the baselines. Then, we report results of experiments on the 30Music open dataset.

\subsection{Setup}
We evaluate the recommendation methods on the next event prediction task. We consider time ordered sequences of user interactions with the items. We split each sequence into the consequent training, validation and test sets. We fit the embedding \textit{Prod2Vec} and \textit{Meta-Prod2Vec} models on the first (n-2) elements of each user sequence and use the performance on the (n-1)-th element to bench the hyper-parameters and we report our final by training on the first (n-1) items of each sequence and predicting the nth item. 
	
We use the last item in the training sequence as the \textit{query item} and we recommend the most similar products using one of the methods described below.

As mentioned in Section \ref{sec:intro}, due to the technical constraint of keeping a constant memory footprint, we are interested in the usefulness of item metadata only at training time. Therefore we do not compare against methods where the metadata is used directly at prediction time, such as the supervised content-based embedding models proposed in \cite{lightfm} where both the user and item are represented as linear combinations of the item content embeddings and \cite{veit2015learning}, where the products are represented by the associated image content embeddings.

We use the following evaluation metrics averaged over all users:
\begin{itemize}
\item Hit ratio at K (HR@K) that is equal to $1/K$ if the test product appears in the top K list of recommended products.
\item Normalized Discounted Cumulative Gain (NDCG@K) favors higher ranks of the test product in the list of recommended products. 
\end{itemize}

Using the aforementioned metrics, we compare the following methods: 
\begin{itemize}
\item \textit{BestOf}: this method retrieves the top products sorted by their popularity. This simulates the  frequently encountered recommendation solution based strictly on popularity. 
\item \textit{CoCounts}: the standard CF method which uses the cosine similarity of vectors of co-occurrences with other items. This method is performing particularly well in cases where the catalog of possible items is small and does not change a lot over time, therefore avoiding item cold-start problems.
\item \textit{Standalone Prod2Vec}: the product recommendation solution based on cosine similarities of the vectorial representation of products obtained by using \textit{Word2Vec} on product sequences. Similarly with other embedding and matrix factorization solutions, the goal of the method is to address the cold-start problem.  
\item \textit{Standalone Meta-Prod2Vec}: our proposed method, which enhances \textit{Prod2Vec} with item side information and uses the resulting product embeddings to compute cosine similarities. As in \textit{Prod2Vec}, the goal of the method is to further address cold-start problems.
\item \textit{Mix(Prod2Vec,CoCounts)}: an ensemble method which returns the top items using a linear combination between the \textit{Prod2Vec} and the \textit{CoCount}-based item pair similarities. The motivation of the ensemble methods that combine embeddings and CF is to leverage the benefits of the two in cold-start and non cold-start regimes.
\begin{eqnarray}
Mix(Prod2Vec,CoCounts) = \nonumber \\ \alpha * Prod2Vec + (1-\alpha) * CoCounts
\label{eq:mix_prod2vec}
\end{eqnarray}
\item \textit{Mix(Meta-Prod2Vec,CoCounts)}: an ensemble method which returns the top items using a linear combination between the \textit{Meta-Prod2Vec} and the \textit{CoCount}-based item pair similarity.
\begin{eqnarray}
Mix(MetaProd2Vec,CoCounts) = \nonumber \\ \alpha * MetaProd2Vec + (1-\alpha) * CoCounts
\label{eq:mix_metaprod2vec}
\end{eqnarray}
\end{itemize}	

\begin{table*}
\centering
\begin{tabular}{l|c|c|c|c} \hline
\textbf{Method} & HR@10 & NDCG@10 & HR@20 & NDCG@20\\ \hline
BestOf & 0.0003 (0.0002;0.0003) & 0.001 (0.001;0.001) & 0.0003 (0.0002;0.0003) & 0.002 (0.002;0.002) \\
CoCounts & 0.0248 (0.0245;0.0251) & 0.122 (0.121;0.123) & 0.0160 (0.0158;0.0161) & 0.141 (0.139;0.142) \\
Prod2Vec & 0.0170 (0.0168;0.0171) & 0.105 (0.103;0.106) & 0.0101 (0.0100;0.0102) & 0.113 (0.112;0.115) \\
Meta-Prod2Vec & 0.0191 (0.0189;0.0194) & 0.110 (0.108;0.113) & 0.0124 (0.0123;0.0126) & 0.125 (0.123;0.126) \\
Mix(Prod2Vec,CoCounts) & 0.0273 (0.027;0.0276) & 0.140 (0.139;0.141) & 0.0158 (0.0157;0.0160) & 0.152 (0.151;0.153) \\
Mix(Meta-Prod2Vec,CoCounts) & \textbf{0.0292} (0.0288;0.0297) & \textbf{0.144} (0.142;0.145) & \textbf{0.0180} (0.0178;0.0182) & \textbf{0.161} (0.160;0.162) \\ 
\hline\end{tabular}
\caption{Comparison of recommendation performance of Meta-Prod2Vec and competing models in terms of HitRate and NDCG.}
\label{tab:reco-accuracy}
\end{table*}

\begin{table}[t]
\centering
\begin{tabular}{l|r|r} \hline
\textbf{Method} & Pair freq=0 & Pair freq<3 \\ \hline
BestOf & 0.0002 & 0.0002 \\
CoCounts & 0.0000 & 0.0197 \\
Prod2Vec & 0.0003 & 0.0078 \\
Meta-Prod2Vec & \textbf{0.0013} & 0.0198 \\
Mix(Prod2Vec,CoCounts) & 0.0002 & 0.0200 \\
Mix(Meta-Prod2Vec,CoCounts) & 0.0007 & \textbf{0.0291} \\
\hline\end{tabular}
\caption{Recommendation accuracy (HR@20) in cold-start regime as a function of training frequency of the pair (\textit{query item}, \textit{next item}).}
\label{tab:reco-cold-start}
\end{table}

\begin{table}[t]
\centering
\begin{tabular}{l|r|r} \hline
\textbf{Side Information} & \multicolumn{2}{c}{\% lift} \\
& HR@20 & NDCG@20 \\ \hline
Meta-Prod2Vec with only I|M & 27\% & 32\% \\
Meta-Prod2Vec with only M|I & 50\% & 52\% \\
Meta-Prod2Vec with only J|M & 55\% & 60\% \\
Meta-Prod2Vec without   M|M & 61\% & 65\%\\
\hline\end{tabular}
\caption{Proportion of the Meta-Prod2Vec lift over BestOf due to each type of side information.}
\label{tab:reco-accuracy-side-info}
\end{table}

\subsection{Dataset}
We perform our evaluation on the publicly available 30Music dataset \cite{dataset30music} that represents a collection of listening and playlists data retrieved from Internet radio stations through Last.fm API. On this dataset, we evaluate the recommendation methods on the task of next song prediction.
For the \textit{Meta-Prod2Vec} algorithm we make use of track metadata, namely the artist information.
We run our experiments on a sample of 100k user sessions of the dataset with the resulting vocabulary size of 433k songs and 67k artists.

\subsection{Results}
We keep the embedding dimension fixed to 50, window size to 3, the side information regularization parameter $\lambda$ to 1. We bench the blending factor $\alpha$ in Equations \ref{eq:mix_prod2vec},\ref{eq:mix_metaprod2vec} of the embedding-based similarity score with the \textit{CoCounts}-based cosine similarity score and find that the best value $\alpha^*=0.15$. In addition, we vary the number of training epochs and we find the best parameter value to be 30 for \textit{Prod2Vec} and 10 for \textit{Meta-Prod2Vec}.
As shown in Table \ref{tab:reco-accuracy}, \textit{Meta-Prod2Vec} has better performance both standalone and in the ensemble model with respect to the \textit{Prod2Vec} model (results computed at 90\% confidence levels).

\subsubsection{Improvements on Cold-Start}
\begin{figure}[t]
	\centering
  \includegraphics[scale=0.32]{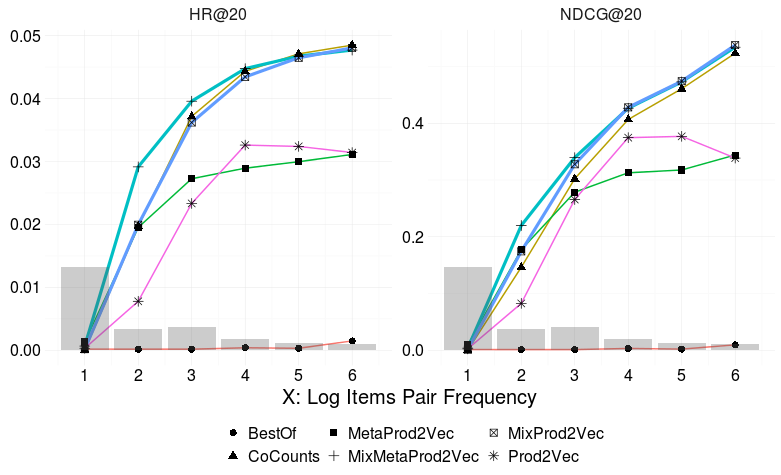}
  \caption{Cold-start improvements on the query and next item pairs.}
  \label{fig:30music_cold-start_log_pair}
\end{figure}

Most of the gains in performance are coming from the cold-start traffic, where, as shown in Figure \ref{fig:30music_cold-start_log_pair} and Table \ref{tab:reco-cold-start}, we can see that \textit{Standalone Meta-Prod2Vec} outperforms by a large margin all other methods when the true pair of (query item, next item) have zero co-occurrences in the training set and that the \textit{Mix(Meta-Prod2Vec,CoCounts)} has the best performance in the cases where the true pair has low observed counts.
Interestingly, we observe that the difference of performance between the ensemble models is bigger than between the standalone \textit{Meta-Prod2Vec} and \textit{Prod2Vec} models.  We explain this by the fact \textit{Standalone Meta-Prod2Vec} outperforms \textit{Standalone Prod2Vec} on the cold-start traffic therefore helping the overall performance of the \textit{CoCounts}, that by itself performs really well on head traffic. Indeed, \textit{CoCounts} memorizes frequent pairs of (query, target) product, while \textit{Standalone Meta-Prod2Vec} helps to generalize on unseen ones. These results are mirrored by similar findings covered in \cite{cheng2016wide} and motivate the newly introduced approach of Wide and Deep learning.

\subsubsection{Relative Importance of each type of Side Information}
We posited the relevance of each type of side information, but we want to confirm experimentally that each of the four types of side-information brings additional information. We proceed by introducing each of the types of the side information separately and to compare its performance with the original \textit{Prod2Vec} baseline. 

The only exception for which we test the value by leaving it out of the full \textit{Meta-Prod2Vec} model is the $mm$ type of side-information; in this case, the input metadata explains the output metadata and this constraint is not directly valuable in regularizing the product embeddings and needs to be introduced together with the other types of pairs that connect the products to the metadata. Therefore its contribution to the model can be computed as (1 - degraded model performance) and is therefore $39\%$ on HR and $35\%$ on NDCG.

In Table \ref{tab:reco-accuracy-side-info}, we compute the proportion of lift over the \textit{BestOf} baseline obtained by using each type of side information separately and we observe that each one of them account for at most 50\% of the performance of the full \textit{Meta-Prod2Vec}, therefore confirming that the additional terms in our proposed model are relevant.

\section{Conclusions}
\label{sec:conclusion}
In this paper, we introduced \textit{Meta-Prod2Vec}, a new item embedding method that enhances the existing \textit{Prod2Vec} method with item metadata at training time. This work makes a novel connection between the  recent embedding-based methods and consecrated Matrix Factorization methods by introducing learning with side information in the context of embeddings. We analyzed separately the relative value of each type of side information and proved that each one of the four types is informative. Finally, we have shown that \textit{Meta-Prod2Vec} constantly outperforms \textit{Prod2Vec} on recommendation tasks both globally and in the cold-start regime, and that, when combined with a standard Collaborative Filtering approach, outperforms all other tested methods. These results, together with the reduced implementation cost and the fact that our method does not affect the online recommendation architecture, makes this solution attractive in cases where item embeddings are already in use.
Future work will extend on ways of using item metadata as side information and support of non-categorical information such as images and continuous variables.

\bibliographystyle{abbrv}

\end{document}